\documentclass[11pt]{article}

\usepackage[preprint]{acl}

 \usepackage{microtype,amsfonts}
\usepackage{booktabs}
\usepackage{multirow}
\usepackage{graphicx}
\usepackage{adjustbox}
\usepackage{xcolor}
\usepackage{colortbl}
\usepackage{tabularx}
\usepackage{array}
\usepackage{amsmath}
\usepackage{pifont}
\definecolor{globalgreen}{HTML}{E8F5E9}
\definecolor{localred}{HTML}{FDECEA}
\definecolor{histblue}{HTML}{EAF2F8}
\definecolor{enhancerorange}{HTML}{FFF3E0}
\definecolor{headergray}{HTML}{ECEFF1}
\definecolor{lightyellow}{HTML}{FFFDE7}

\newcommand{\bestcell}[1]{\cellcolor{lightyellow}\textbf{#1}}
\newcommand{\taskglobal}[1]{\cellcolor{globalgreen}#1}
\newcommand{\tasklocal}[1]{\cellcolor{localred}#1}
\newcommand{\taskhistone}[1]{\cellcolor{histblue}#1}
\newcommand{\taskenhancer}[1]{\cellcolor{enhancerorange}#1}
\usepackage[english,bidi=default]{babel} 
\usepackage{booktabs}
\usepackage{graphicx}

\usepackage{multicol}
\usepackage{multirow}
\usepackage{url}
\usepackage{hyperref}

\title{Frozen but Not Always Accessible: A Representation Analysis of Genomic Language Models}

\author{
Nirjhor Datta$^{1,2}$ \quad
Swakkhar Shatabda$^{2}$ \quad
M. Sohel Rahman$^{1}$ \\[0.4em]
\begin{tabular}{c}
\footnotesize $^{1}$Department of Computer Science and Engineering, Bangladesh University of Engineering and Technology, \\
\footnotesize West Palashi, Dhaka 1205, Bangladesh \\
\footnotesize $^{2}$Department of Computer Science and Engineering, BRAC University, Dhaka 1205, Bangladesh
\end{tabular}
}
\begin{document}
\maketitle
\begin{abstract}
Genomic foundation models are increasingly reused as frozen feature extractors for downstream sequence prediction, offering a compute-efficient alternative to full fine-tuning. However, it remains unclear when biological information encoded by these models is accessible without task-specific adaptation. We present a representation-accessibility analysis of frozen genomic language models across regulatory, epigenetic, promoter, splice-site, and variant-effect prediction tasks. We evaluate DNABERT-2, Nucleotide Transformer, HyenaDNA, GENERATOR-v2, and Omni-DNA under unified frozen-probing protocols, while separating diagnostic readout analyses from validation-selected checks. Our results reveal a consistent task-dependent pattern: frozen probes recover 95--100\% of fine-tuned performance on promoter tasks, but average splice-site recovery drops to 60--88\%. Frozen embeddings are also competitive on broad Genomic Benchmark tasks such as coding-region and species-discrimination classification, but show larger gaps on some regulatory and OCR tasks. Layer-wise probing, in-silico mutagenesis, variant-effect prediction, and embedding geometry show that local biological signal is partially present in frozen representations, but is not always accessible through final pooled embeddings.
\end{abstract}

\section{Introduction}

Genomic foundation models are increasingly used as pretrained encoders for DNA sequence analysis. Models such as DNABERT~\citep{ji2021dnabert}, DNABERT-2~\citep{zhou2023dnabert}, Nucleotide Transformer~\citep{dalla2025nucleotide}, HyenaDNA~\citep{nguyen2023hyenadna}, Caduceus~\citep{schiff2024caduceus}, GENERATOR~\citep{wu2025generator}, and Omni-DNA~\citep{li2025omni} have shown strong promise for regulatory prediction, sequence classification, and variant interpretation. Related biological sequence modeling studies have also appeared in NLP venues, including genomic and proteomic multimodal models~\citep{liu2024geneverse}, RNA foundation models~\citep{yang2024mp}, multi-omics instruction benchmarks~\citep{he2024biology}, and RNA-binding prediction frameworks~\citep{jiang2025rbptool}. These developments raise a practical question: when can frozen genomic representations replace task-specific fine-tuning?

Freezing a pretrained encoder and training a lightweight classifier on extracted embeddings is computationally attractive. It avoids expensive fine-tuning, reduces GPU requirements, and allows embeddings to be reused across tasks. However, strong fine-tuned performance does not imply that all task-relevant biological information is easily accessible from frozen representations. A model may encode useful sequence information, but that signal may be distributed across token states, concentrated in intermediate layers, diluted by pooling, or poorly separated in the final embedding space. Thus, the central question is not only whether genomic foundation models contain biological information, but whether that information is accessible to simple downstream probes.

This distinction is important because genomic tasks require different biological signals. Promoter classification, coding-region classification, and species discrimination often depend on global composition or distributed sequence patterns. In contrast, splice-site and variant-effect prediction require sensitivity to precise local nucleotide changes and position-specific mechanisms. A final pooled embedding may be sufficient for global tasks while failing to expose local mechanistic signals. The Nucleotide Transformer benchmark evaluates genomic
foundation models across enhancer, histone-mark, promoter, and splice-site
tasks~\citep{dalla2025nucleotide}, while Genomic Benchmark and GenBench
provide standardized sequence-classification settings~\citep{grevsova2023genomic,liu2024genbench}.
For variant-level evaluation, long-range genomic benchmarks include tasks such
as Causal eQTL and Pathogenic ClinVar~\citep{trop2024genomics}. However, these
resources do not by themselves answer when task-relevant biological information
is accessible from frozen representations under lightweight probing.

In this work, we study \emph{representation sufficiency} and \emph{representation accessibility} in frozen genomic language models. We evaluate DNABERT-2, Nucleotide Transformer, HyenaDNA, GENERATOR-v2, and Omni-DNA across regulatory, epigenetic, promoter, splice-site, and variant-effect prediction tasks. These models cover complementary families, including transformer encoders, long-sequence Hyena-based models, large genomic generative models adapted for embedding extraction, and unified genomic representation models. Our goal is not to exhaustively benchmark all genomic foundation models, but to test whether accessibility patterns persist across model families and task types. We treat NT-v3 and Evo~2 as important scope boundaries because their primary interfaces emphasize long-context sequence-function modeling, functional-track prediction, generation, or zero-shot/embedding-based variant analysis rather than the short sequence-level frozen classification protocol studied here.

Our results show a consistent task-dependent pattern: frozen representations recover most fine-tuned performance on promoter and other global or composition-driven tasks, but recover substantially less on splice-site and variant-effect tasks. The same pattern is observed across both established backbones and recent models such as GENERATOR-v2 and Omni-DNA. We further show that readout choice matters, but does not fully explain the local-task gap: diagnostic grid-best analyses reveal readout sensitivity, while validation-selected checks preserve the same qualitative conclusion. Layer-wise probing, in-silico mutagenesis, variant-effect analysis, and embedding geometry indicate that local biological signal is partially present in token-level or intermediate representations, but is not always accessible through final pooled embeddings.

Our study makes five contributions, organized around six research questions.
First, we analyze when frozen genomic language-model representations are sufficient across regulatory, epigenetic, promoter, splice-site, and variant-effect prediction tasks (\textbf{RQ1}). 
Second, we evaluate multiple genomic foundation-model families, including DNABERT-2, Nucleotide Transformer, HyenaDNA, GENERATOR-v2, and Omni-DNA, and show that the promoter-versus-splice accessibility gap is consistent across model families (\textbf{RQ1}). 
Third, we study how pooling and readout choices affect accessibility by comparing mean, max, CLS, decoder last-token, norm-attention, and top-$k$ pooling, while separating diagnostic grid-best analyses from validation-selected checks (\textbf{RQ2}). 
Fourth, we diagnose local biological signal using layer-wise probing, in-silico mutagenesis, variant-effect representation shifts, and quantitative embedding geometry (\textbf{RQ3--RQ5}). 
Finally, we compare frozen probing with controlled DNABERT-2 LoRA fine-tuning under matched preprocessing, splits, metrics, and readouts, while marking externally reported fine-tuned scores as reference values for broad recovery analysis (\textbf{RQ6}).

\section{Method}
\begin{figure*}[t]
    \centering    \includegraphics[width=\textwidth]{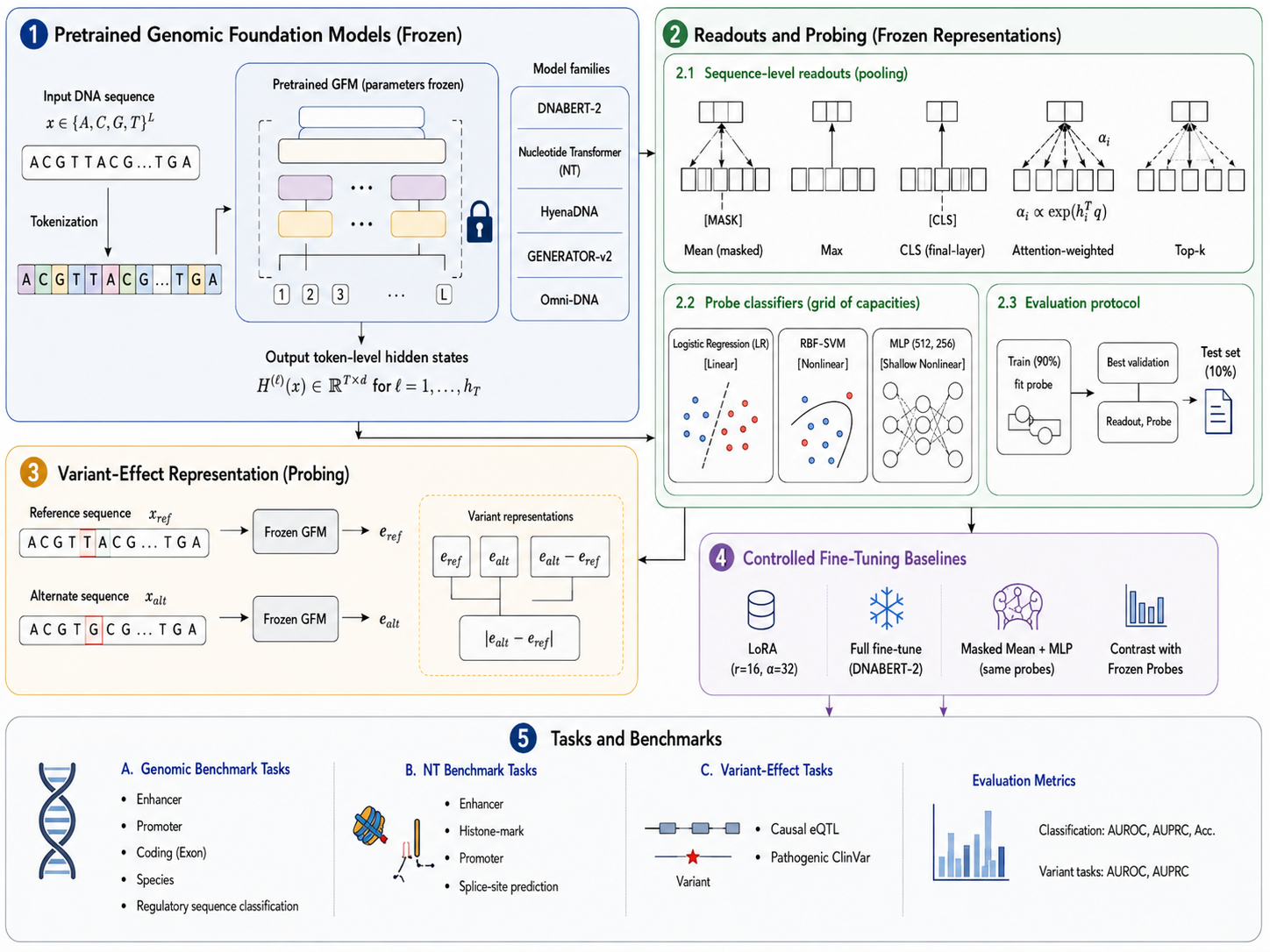}
    \caption{Top-level overview of the proposed framework.}
    \label{fig:overview}
\end{figure*}

Detailed description of the proposed methods is presented at Figure \ref{fig:overview}.
\subsection{Task Definition and Representation Sufficiency}

We study genome sequence classification. Each example consists of a DNA sequence
$x=(x_1,\ldots,x_L)$, where $x_i\in\{A,C,G,T\}$, and a label $y$.
For binary tasks, $y\in\{0,1\}$; for multi-class tasks,
$y\in\{1,\ldots,K\}$. Our goal is to quantify
\emph{representation sufficiency}: the extent to which downstream biological
signal is accessible from frozen genomic language-model representations using
lightweight decoders. We use the gap between frozen-probe performance and
fine-tuned performance as an indicator of when task-specific adaptation provides
additional benefit.

Let $f_\theta$ denote a pretrained genomic language model with frozen parameters
$\theta$. Given an input sequence $x$, the model produces token-level hidden
states
\begin{equation}
\footnotesize
H^{(\ell)}(x)=\left(h^{(\ell)}_1,\ldots,h^{(\ell)}_T\right),
\end{equation}
where $\ell$ denotes the layer, $T$ is the token length, and
$h^{(\ell)}_i\in\mathbb{R}^d$. Unless otherwise specified, we use the final-layer
hidden states and write them as $h_i$ for simplicity.

\subsection{Frozen Backbones}

We evaluate five genomic foundation-model families with different architectures
and training regimes:
\begin{itemize}
\itemsep0em 
    \item DNABERT-2~\citep{zhou2024dnabert}-117M;
    \item Nucleotide Transformer (NT)~\citep{dalla2025nucleotide}-500M;
    \item HyenaDNA~\citep{nguyen2023hyenadna}-436K;
    \item GENERATOR-v2~\citep{wu2025generator}-1.2B;
    \item Omni-DNA~\citep{li2025omni}-116M.
\end{itemize}
All frozen-representation experiments use the backbone strictly in inference
mode: no model parameters are updated, gradients are disabled, and only the
extracted sequence representations are used for downstream probing.

\subsection{Pooling and Readout Functions}

We evaluate multiple sequence-level readouts from token-level hidden states to
test whether task-relevant information is lost or diluted during pooling. Let
$m_i\in\{0,1\}$ denote the attention mask, and let
$\mathcal{I}=\{i:m_i=1\}$ be the set of valid non-padding tokens.

\paragraph{Mean pooling.}
\begin{equation}
\footnotesize
e_{\mathrm{mean}}(x)=
\frac{\sum_{i\in\mathcal{I}} h_i}{|\mathcal{I}|}.
\end{equation}

\paragraph{Max pooling.}
\begin{equation}
\footnotesize
e_{\mathrm{max}}(x)=\max_{i\in\mathcal{I}} h_i.
\end{equation}

\paragraph{CLS pooling.}
For encoder models with a CLS token, we use the final-layer CLS representation
as the sequence embedding.

\paragraph{Last-token pooling.}
For decoder-style models, we use the last non-padding token representation as
the sequence embedding.

\paragraph{Norm-attention pooling.}
Token states are weighted according to their representation norm:
\begin{equation}
\footnotesize
\alpha_i=
\frac{e^{\|h_i\|_2}}{\sum_{j\in\mathcal{I}}e^{\|h_j\|_2}},
\quad
e_{\mathrm{attn}}(x)=\sum_{i\in\mathcal{I}}\alpha_i h_i .
\end{equation}

\paragraph{Top-$k$ pooling.}
We select the $k$ valid token states with the largest representation norms and
average them to obtain the sequence embedding.

\subsection{Probe Classifiers}

To evaluate accessibility of task-relevant information, we train lightweight
probes on frozen embeddings. We use:
\begin{itemize}
\itemsep0em 
    \item Logistic Regression (LR), measuring linear accessibility;
    \item RBF-SVM, measuring nonlinear kernel decodability;
    \item MLP, a shallow nonlinear classifier with hidden layers $(512,256)$.
\end{itemize}
Embeddings are standardized using training-set statistics before probe training.
No validation or test statistics are used during standardization.

\subsection{Readout Selection and Diagnostic Grid}

We evaluate the grid

$\mathcal{G}=\{\mathrm{mean},\mathrm{max},\mathrm{CLS},\mathrm{last},
\mathrm{attn},\mathrm{top}\text{-}k\}\times\{\mathrm{LR},\mathrm{MLP}\}$,
with architecture-specific availability of CLS and last-token readouts.

We distinguish two uses of this grid. First, we report diagnostic grid-best
results to measure how much performance can vary as the readout changes. These
results are used as readout-sensitivity analyses and are not interpreted as
unbiased model-selection estimates. Second, where validation experiments are
available, we perform validation-selected readout analysis. For this setting, we
hold out 10\% of the training set as a stratified validation split, train each
configuration on the remaining 90\%, select the configuration with the highest
mean validation score across seeds, and report its test performance. The test
set is used only for final reporting and never for readout selection.

\subsection{Variant-Effect Representation}

For variant-effect prediction, each example consists of a reference sequence
$x_{\mathrm{ref}}$ and an alternative sequence $x_{\mathrm{alt}}$. We compute
frozen embeddings independently:
\begin{equation}
\footnotesize
e_{\mathrm{ref}}=e(x_{\mathrm{ref}}), \qquad
e_{\mathrm{alt}}=e(x_{\mathrm{alt}}).
\end{equation}
We represent the variant using the concatenated feature vector
\begin{equation}
\footnotesize
\phi(x_{\mathrm{ref}},x_{\mathrm{alt}})
=
\left[
e_{\mathrm{ref}},
e_{\mathrm{alt}},
e_{\mathrm{alt}}-e_{\mathrm{ref}},
|e_{\mathrm{alt}}-e_{\mathrm{ref}}|
\right].
\end{equation}
This representation captures both the original sequence context and the
direction and magnitude of mutation-induced representation change.

\subsection{Controlled Fine-Tuning Baseline}

For a matched frozen-versus-adapted comparison, we reproduce DNABERT-2 LoRA
fine-tuning under the same setup used for frozen probing. The adapted and frozen
settings use the same DNABERT-2 checkpoint, the same maximum sequence length,
the same preprocessing code path, the same train/test split loader, the same
metric functions, and the same masked mean-pooling plus MLP classification head.
Thus, the primary variable is whether encoder weights are adapted.

Fine-tuning uses LoRA with rank $r=16$, $\alpha=32$, dropout 0.05, AdamW, batch
size 32, learning rate $5\times10^{-5}$, and a cosine schedule. We use this as a
conservative adapted reference rather than a fully tuned upper bound. Other
fine-tuned scores used in broad recovery tables are benchmark- or model-reported
where available and are therefore treated as provenance-marked reference values
rather than fully matched head-to-head comparisons.
\section{Experimental Setup}

\subsection{Datasets}

We evaluate on three groups of tasks.

\paragraph{Genomic Benchmark.}
We use Genomic Benchmark tasks covering enhancer, promoter, coding-region,
species-discrimination, regulatory, and OCR sequence classification
settings~\citep{grevsova2023genomic,liu2024genbench}. These tasks evaluate
whether frozen genomic representations capture broad global or
composition-driven sequence features.

\paragraph{Nucleotide Transformer Benchmark.}
We use the Nucleotide Transformer benchmark~\citep{dalla2025nucleotide},
which includes enhancer, histone-mark, promoter, and splice-site prediction
tasks. These tasks allow us to compare global or composition-driven signals,
such as promoter prediction, with local and position-sensitive biological
signals, such as splice-site prediction.

\paragraph{Variant-effect prediction.}
We use Causal eQTL and Pathogenic ClinVar from the Genomics Long-Range
Benchmark~\citep{trop2024genomics}. These tasks evaluate whether frozen
representations capture functional changes induced by reference-to-alternative
sequence variants.

\subsection{Metrics}

We use the standard metric for each benchmark family. MCC is used for enhancer
and histone-mark tasks, F1-score for promoter and splice-site tasks, and the
benchmark-reported metric is used for variant-effect tasks where applicable. To
compare frozen probing with fine-tuned references, we compute recovery as
\begin{equation}
\footnotesize
\mathrm{Recovery}=
\frac{\mathrm{Frozen\ Probe\ Score}}
{\mathrm{Fine\ Tuned\ Score}}
\times 100.
\end{equation}

\subsection{Reporting}

Unless otherwise specified, frozen-probe results are reported as
mean$\pm$standard deviation over five seeds. Logistic regression is deterministic
given fixed embeddings and splits, while MLP probes vary due to random
initialization and optimization. Diagnostic grid-best results are reported as
readout-sensitivity analyses. Validation-selected results are reported
separately when available, using the held-out validation protocol described
above. When fine-tuned scores are taken from benchmark or model-reported values
rather than reproduced under our protocol, we mark them as reference fine-tuned
scores and do not interpret them as fully controlled head-to-head comparisons.

\section{Results and  Analysis}

\noindent\textbf{\textit{RQ1: When Are Frozen Representations Sufficient Across Task Families?}}
RQ1 asks whether frozen representations are equally sufficient across biological
task families. We begin with the NT benchmark because it directly contrasts
global promoter tasks with local splice-site tasks, then use Genomic Benchmark
to test broader binary sequence-classification settings.

\noindent\textbf{\textit{Task-Family Analysis on the NT Benchmark.}} Table~\ref{tab:main_nt_multimodel} reports frozen-probe recovery across
HyenaDNA, NT, GENERATOR-v2, and Omni-DNA. Fine-tuned values are benchmark- or
model-reported where available, while frozen-probe scores use our unified
extraction and probing protocol.

\begin{table*}[t]
\centering
\scriptsize
\setlength{\tabcolsep}{1.6pt}
\renewcommand{\arraystretch}{0.96}
\caption{Main NT benchmark evaluation. Each cell reports Frozen / FT / Recovery. FT values are benchmark/model-reported where available; frozen values use our unified probing protocol.}
\label{tab:main_nt_multimodel}
\resizebox{\textwidth}{!}{%
\begin{tabular}{lcccc}
\toprule
\rowcolor{headergray}
\textbf{Task} & \textbf{HyenaDNA} & \textbf{NT} & \textbf{GENERATOR-v2} & \textbf{Omni-DNA} \\
\midrule
\taskhistone{H3} & .6926$\pm$.0158 / .779$\pm$.037 / 89 & .6952$\pm$.0111 / .784$\pm$.047 / 89 & .7381$\pm$.0150 / .806 / 92 & .7220$\pm$.0000 / .818$\pm$.005 / 88 \\
\taskhistone{H3K14ac} & .3295$\pm$.0132 / .612$\pm$.065 / 54 & .3315$\pm$.0068 / .551$\pm$.021 / 60 & .5036$\pm$.0046 / .605 / 83 & .3873$\pm$.0052 / .685$\pm$.014 / 57 \\
\taskhistone{H3K36me3} & .3961$\pm$.0095 / .613$\pm$.041 / 65 & .3999$\pm$.0160 / .625$\pm$.013 / 64 & .5667$\pm$.0000 / .657 / 86 & .4739$\pm$.0000 / .661$\pm$.013 / 72 \\
\taskhistone{H3K4me1} & .3315$\pm$.0107 / .512$\pm$.024 / 65 & .3278$\pm$.0152 / .550$\pm$.021 / 60 & .4997$\pm$.0134 / .553 / 90 & .3790$\pm$.0000 / .577$\pm$.083 / 66 \\
\taskhistone{H3K4me2} & .2792$\pm$.0079 / .455$\pm$.095 / 61 & .2535$\pm$.0128 / .319$\pm$.045 / 79 & .3159$\pm$.0000 / .424 / 75 & .2658$\pm$.0143 / .576$\pm$.003 / 46 \\
\taskhistone{H3K4me3} & .2215$\pm$.0100 / .549$\pm$.056 / 40 & .2037$\pm$.0227 / .410$\pm$.033 / 50 & .3822$\pm$.0000 / .512 / 75 & .2512$\pm$.0000 / .587$\pm$.222 / 43 \\
\taskhistone{H3K79me3} & .5499$\pm$.0085 / .672$\pm$.048 / 82 & .5660$\pm$.0085 / .672$\pm$.048 / 84 & .5621$\pm$.0000 / .670 / 84 & .5621$\pm$.0000 / .718$\pm$.027 / 78 \\
\taskhistone{H3K9ac} & .4299$\pm$.0000 / .581$\pm$.061 / 74 & .3994$\pm$.0155 / .562$\pm$.040 / 71 & .6098$\pm$.0000 / .612 / 100 & .4545$\pm$.0000 / .658$\pm$.029 / 69 \\
\taskhistone{H4} & .7267$\pm$.0000 / .763$\pm$.044 / 95 & .7144$\pm$.0084 / .799$\pm$.025 / 89 & .7928$\pm$.0078 / .815 / 97 & .7501$\pm$.0000 / .802$\pm$.002 / 94 \\
\taskhistone{H4ac} & .3041$\pm$.0171 / .564$\pm$.038 / 54 & .3025$\pm$.0000 / .495$\pm$.032 / 61 & .4439$\pm$.0035 / .592 / 75 & .3222$\pm$.0167 / .663$\pm$.029 / 49 \\
\midrule
\taskenhancer{Enhancer} & .5243$\pm$.0635 / .517$\pm$.117 / 101 & .5252$\pm$.0262 / .548$\pm$.144 / 96 & .5314$\pm$.0238 / .580 / 92 & .4947$\pm$.0276 / .593$\pm$.005 / 83 \\
\taskenhancer{Enhancer Types} & .3710$\pm$.0187 / .386$\pm$.185 / 96 & .5252$\pm$.0262 / .424$\pm$.132 / 124 & .4305$\pm$.0000 / .477 / 90 & .4129$\pm$.0345 / .498$\pm$.001 / 83 \\
\midrule
\taskglobal{Promoter All} & .9329$\pm$.0004 / .960$\pm$.005 / 97 & .9279$\pm$.0042 / .976$\pm$.006 / 95 & .9630$\pm$.0000 / .962 / 100 & .9524$\pm$.0000 / .973$\pm$.002 / 98 \\
\taskglobal{Promoter Non-TATA} & .9350$\pm$.0016 / .959$\pm$.008 / 97 & .9373$\pm$.0024 / .976$\pm$.005 / 96 & .9662$\pm$.0000 / .962 / 100 & .9539$\pm$.0000 / .972$\pm$.016 / 98 \\
\taskglobal{Promoter TATA} & .9338$\pm$.0000 / .944$\pm$.040 / 99 & .9182$\pm$.0056 / .966$\pm$.013 / 95 & .9525$\pm$.0080 / .948 / 100 & .9539$\pm$.0000 / .967$\pm$.001 / 99 \\
\midrule
\tasklocal{Splice All} & .5243$\pm$.0051 / .956$\pm$.011 / \bestcell{55} & .6871$\pm$.0083 / .983$\pm$.008 / \bestcell{70} & .7754$\pm$.0000 / .978 / \bestcell{79} & .4221$\pm$.0000 / .927$\pm$.025 / \bestcell{46} \\
\tasklocal{Splice Acceptor} & .4752$\pm$.0108 / .958$\pm$.010 / 50 & .8600$\pm$.0012 / .981$\pm$.011 / 88 & .9089$\pm$.0000 / .981 / 93 & .8147$\pm$.0049 / .968$\pm$.002 / 84 \\
\tasklocal{Splice Donor} & .7247$\pm$.0087 / .949$\pm$.024 / 76 & .8099$\pm$.0054 / .985$\pm$.022 / 82 & .9094$\pm$.0021 / .978 / 93 & .8082$\pm$.0054 / .951$\pm$.007 / 85 \\
\midrule
\rowcolor{globalgreen}
\textbf{Promoter avg. recovery} & \textbf{98} & \textbf{95} & \textbf{100} & \textbf{98} \\
\rowcolor{localred}
\textbf{Splice avg. recovery} & \textbf{60} & \textbf{80} & \textbf{88} & \textbf{72} \\
\bottomrule
\end{tabular}%
}
\end{table*}

The NT benchmark shows a clear task-dependent accessibility pattern. Promoter
tasks recover 98\%, 95\%, 100\%, and 98\% of fine-tuned performance for
HyenaDNA, NT, GENERATOR-v2, and Omni-DNA, respectively. In contrast, splice
recovery drops to 60\%, 80\%, 88\%, and 72\%, with \textit{Splice All} reaching
only 55\%, 70\%, 79\%, and 46\%. Thus, the promoter-versus-splice accessibility
gap is consistent across both established and recent model families.

\noindent\textbf{\textit{Broad Binary Sequence Classification on Genomic Benchmark.}} The Genomic Benchmark results provide broader support for this pattern.
Table~\ref{tab:genbench_main} shows that frozen embeddings are strong on several
broad binary tasks, including coding-region and species-discrimination
classification, but still trail fine-tuning on average.

\begin{table*}[t]
\centering
\scriptsize
\setlength{\tabcolsep}{2.6pt}
\renewcommand{\arraystretch}{0.92}
\caption{Genomic Benchmark results. FT denotes fine-tuning; Emb(best) denotes the best frozen-embedding probe.}
\label{tab:genbench_main}
\resizebox{\textwidth}{!}{%
\begin{tabular}{lcccccc}
\toprule
\rowcolor{headergray}
\multirow{2}{*}{\textbf{Dataset}} 
& \multicolumn{2}{c}{\textbf{DNABERT-2}} 
& \multicolumn{2}{c}{\textbf{NT}} 
& \multicolumn{2}{c}{\textbf{HyenaDNA}} \\
\cmidrule(lr){2-3} \cmidrule(lr){4-5} \cmidrule(lr){6-7}
\rowcolor{headergray}
& \textbf{FT} & \textbf{Emb(best)}
& \textbf{FT} & \textbf{Emb(best)}
& \textbf{FT} & \textbf{Emb(best)} \\
\midrule
\taskenhancer{Mouse Enhancers}              & 81.82 & 76.03 & 85.12 & 80.58 & 79.34 & 73.14 \\
\taskglobal{Coding vs Intergenic}           & 93.58 & 93.35 & 95.76 & 89.42 & 90.97 & 88.32 \\
\taskglobal{Human vs Worm}                  & 97.39 & 96.80 & 97.51 & 93.48 & 96.24 & 93.79 \\
\taskenhancer{Human Enhancers (Cohn)}       & 75.87 & \bestcell{77.00} & 76.12 & 73.89 & 72.96 & 72.87 \\
\taskenhancer{Human Enhancers (Ensembl)}    & 90.75 & 78.86 & 92.44 & 73.35 & 90.33 & 72.92 \\
\taskenhancer{Human Ensembl Regulatory}     & 88.32 & 86.19 & 94.03 & 90.61 & 84.62 & \bestcell{84.90} \\
\taskglobal{Human Non-TATA Promoters}       & 95.24 & 93.71 & 96.60 & 84.54 & 94.45 & 86.21 \\
\taskenhancer{Human OCR (Ensembl)}          & 75.82 & 69.60 & 80.42 & \bestcell{90.61} & 79.14 & 68.93 \\
\midrule
\rowcolor{headergray}
\textbf{Average} 
& \textbf{87.97} & \textbf{81.49} 
& \textbf{89.75} & \textbf{85.49} 
& \textbf{85.50} & \textbf{79.39} \\
\bottomrule
\end{tabular}%
}
\end{table*}

On Genomic Benchmark, frozen embeddings nearly match fine-tuning for
DNABERT-2 on \textit{Coding vs Intergenic} and \textit{Human vs Worm}, and
slightly exceed fine-tuning on \textit{Human Enhancers (Cohn)}. However, average
frozen-probe performance remains below fine-tuning for all three backbones.
Together with the NT benchmark, these results show that frozen representations
are useful for many global sequence-classification tasks but less sufficient for
local, position-sensitive biological mechanisms.

\noindent\textbf{\textit{RQ2: Does Readout Choice Affect Representation Accessibility?}} A possible explanation for weak frozen-probe performance on splice tasks is that
mean pooling dilutes local motif information. We therefore evaluate multiple
readouts, including mean, max, CLS, decoder last-token, norm-attention, and
top-$k$ pooling. Table~\ref{tab:readout_best} is used as a diagnostic
readout-sensitivity analysis: it shows how the best observed configuration
varies across models and tasks, but it is not used as an unbiased
model-selection estimate.

\begin{table*}[t]
\centering
\small
\setlength{\tabcolsep}{6pt}
\caption{Diagnostic best pooling+classifier configuration on representative tasks. The table summarizes readout sensitivity and should not be interpreted as an unbiased validation-selected estimate.}
\label{tab:readout_best}
\rowcolors{2}{white}{gray!6}
\begin{tabular}{lcccc}
\toprule
\rowcolor{headergray}
\textbf{Task} & \textbf{HyenaDNA} & \textbf{NT} & \textbf{GENERATOR-v2} & \textbf{Omni-DNA} \\
\midrule
H3 & MLP+Attn & MLP+CLS & MLP+Mean & LR+Mean \\
H4ac & MLP+Attn & LR+CLS & MLP+Mean & LR+Mean \\
Enhancer & MLP+Attn & MLP+Mean & MLP+Max & MLP+Mean \\
Promoter All & MLP+Mean & MLP+CLS & LR+Mean & MLP+Mean \\
Splice All & MLP+Mean & MLP+CLS & LR+Mean & LR+Mean \\
Splice Acceptor & MLP+Attn & MLP+CLS & LR+Mean & MLP+Mean \\
Splice Donor & MLP+Attn & MLP+CLS & MLP+Mean & LR+Mean \\
\midrule
Non-mean wins & 13/18 & 14/18 & 2/18 & Mean only \\
\bottomrule
\end{tabular}
\end{table*}

Readout choice substantially affects the accessibility of frozen representations.
For HyenaDNA and NT, non-mean readouts often outperform mean pooling, indicating
that task-relevant information can be partially hidden by a poor readout.
However, this table is diagnostic: it measures readout sensitivity and does not
serve as the main unbiased model-selection estimate.

To test whether diagnostic grid-best selection materially changes the
conclusion, we additionally perform validation-selected readout analysis for
HyenaDNA. For each task, 10\% of the training set is held out as a stratified
validation split, shared across configurations. We search five readouts
(mean, max, attention, top-$k1$, top-$k5$) and two probes (LR, MLP), select the
configuration with the highest mean validation score across five seeds, and
report its test performance.

\begin{table*}[t]
\centering
\small
\setlength{\tabcolsep}{5pt}
\renewcommand{\arraystretch}{1.05}
\caption{Validation-selected versus diagnostic test grid-best readout results for HyenaDNA. Validation-selected configurations are chosen using a held-out 10\% stratified validation split and evaluated once on the test set. Diagnostic grid-best results are maxima over the test grid and are reported only as readout-sensitivity analyses.}
\label{tab:validation_readout_hyena}
\resizebox{\textwidth}{!}{%
\begin{tabular}{lcccc}
\toprule
\rowcolor{headergray}
\textbf{Dataset} & \textbf{Val-selected test} & \textbf{Val config} & \textbf{Diagnostic grid-best} & \textbf{Grid-best config} \\
\midrule
H3 & 0.6844 $\pm$ 0.0074 & MLP+Attn & 0.6926 $\pm$ 0.0158 & MLP+Attn \\
H4ac & 0.3122 $\pm$ 0.0072 & MLP+Attn & 0.3041 $\pm$ 0.0171 & MLP+Attn \\
Enhancer & 0.4785 $\pm$ 0.0171 & MLP+Attn & 0.5243 $\pm$ 0.0635 & MLP+Attn \\
Promoter All & 0.9322 $\pm$ 0.0015 & MLP+Mean & 0.9329 $\pm$ 0.0004 & MLP+Mean \\
Splice All & 0.5288 $\pm$ 0.0044 & MLP+Mean & 0.5243 $\pm$ 0.0051 & MLP+Attn \\
Splice Acceptor & 0.7308 $\pm$ 0.0033 & MLP+Mean & 0.7360 $\pm$ 0.0052 & MLP+Attn \\
Splice Donor & 0.7247 $\pm$ 0.0087 & MLP+Mean & 0.7234 $\pm$ 0.0135 & MLP+Attn \\
\bottomrule
\end{tabular}%
}
\end{table*}

Validation-selected and diagnostic grid-best scores agree closely on most tasks.
For example, Promoter All changes from 0.9329 to 0.9322, and Splice All changes
from 0.5243 to 0.5288. The selected configuration is identical on four of seven
tasks. Where configurations differ, the top readouts are often statistically
tied: on Splice All, attention+MLP obtains 0.5243$\pm$0.0051, while mean+MLP
obtains 0.5240$\pm$0.0056, a difference of only 0.0003. We therefore avoid
over-interpreting the identity of the single best readout.

The robust conclusion is that readout choice matters, but the splice gap is not
merely a mean-pooling artifact. Even under stronger readouts, splice tasks
remain much less recoverable than promoter tasks. This supports the broader
interpretation that local biological signals are partially present in frozen
representations, but are not always reliably accessible through simple
sequence-level readouts.

\noindent\textbf{\textit{RQ3: Where Is Local Signal Most Accessible?}}

To test where splice-relevant information is most accessible within the frozen
encoder, we perform layer-wise probing on the NT model for \textit{Splice All}.
Table~\ref{tab:layerwise_nt} reports representative layer-wise results.

\begin{table}[t]
\centering
\scriptsize
\setlength{\tabcolsep}{2.2pt}
\renewcommand{\arraystretch}{0.95}
\caption{Layer-wise probing on NT for \textit{Splice All}. MLP results are averaged over five seeds.}
\label{tab:layerwise_nt}
\resizebox{\columnwidth}{!}{%
\begin{tabular}{lccc}
\toprule
\rowcolor{headergray}
\textbf{Layer} & \textbf{ACC} & \textbf{F1} & \textbf{MCC} \\
\midrule
20 (best) & \bestcell{0.639$\pm$.022} & \bestcell{0.527$\pm$.028} & \bestcell{0.449$\pm$.029} \\
11 & 0.625$\pm$.019 & 0.507$\pm$.013 & 0.441$\pm$.032 \\
24 (final) & $\sim$0.60 & $\sim$0.52 & $\sim$0.36--0.47 \\
\bottomrule
\end{tabular}%
}
\end{table}

\paragraph{Analysis.}
Intermediate layers expose splice-relevant information better than the final-layer
readout. This suggests that local biological signal is not completely absent
from frozen encoders. Instead, some of the signal becomes less accessible after
final-layer aggregation or through the default sequence-level readout. This
supports a more precise interpretation: frozen genomic models partially encode
local splice information, but task-specific adaptation or better representation
extraction is needed to make that signal reliably usable.

\noindent\textbf{\textit{RQ4: Do Frozen Representations Capture Local Perturbation Effects?}} We next evaluate local sensitivity using in-silico mutagenesis. For each held-out
sequence, each nucleotide position is mutated to the three alternative bases,
and we measure the resulting change in the frozen-probe prediction score. This
directly tests whether frozen readouts respond to local nucleotide changes.

\begin{table}[t]
\centering
\scriptsize
\setlength{\tabcolsep}{2.5pt}
\renewcommand{\arraystretch}{0.95}
\caption{In-silico mutagenesis on \textit{splice\_sites\_all}. Top-5 pooling gives stronger position-specific sensitivity than mean pooling.}
\label{tab:ism}
{%
\begin{tabular}{lccc}
\toprule
\rowcolor{headergray}
\textbf{Model} & \textbf{Mean} & \textbf{Top-5} & \textbf{Ratio} \\
\midrule
DNABERT-2 & 0.0417 & 0.0519 & 1.24$\times$ \\
NT & 0.0576 & 0.0714 & 1.24$\times$ \\
\bottomrule
\end{tabular}%
}
\end{table}

Token-aware pooling increases positional sensitivity by 1.24$\times$ for both
DNABERT-2 and NT. This indicates that local signal is partly present in
token-level frozen representations. However, the improvement in sensitivity
does not fully close the splice classification gap, showing that local
perturbation information is only partially accessible to lightweight frozen
probes.

\noindent\textbf{\textit{Variant-Effect Prediction.}} Variant-effect prediction provides a complementary test of local biological
sensitivity because the label depends on the functional impact of a
reference-to-alternative sequence change. Table~\ref{tab:variant_effect}
compares frozen embedding-based approaches with fine-tuned baselines.

\begin{table}[t]
\centering
\scriptsize
\setlength{\tabcolsep}{2.2pt}
\renewcommand{\arraystretch}{0.95}
\caption{Variant-effect prediction on LRB tasks.}
\label{tab:variant_effect}
\resizebox{\columnwidth}{!}{%
\begin{tabular}{lcccccc}
\toprule
\rowcolor{headergray}
\multirow{2}{*}{\textbf{Task}} 
& \multicolumn{3}{c}{\textbf{FT}} 
& \multicolumn{3}{c}{\textbf{Emb.}} \\
\cmidrule(lr){2-4}\cmidrule(lr){5-7}
& \textbf{NT} & \textbf{Hyena} & \textbf{DNA2}
& \textbf{NT} & \textbf{Hyena} & \textbf{DNA2} \\
\midrule
Causal eQTL & 0.72 & 0.72 & 0.72 & 0.70 & 0.66 & 0.69 \\
Path. ClinVar & 0.78 & 0.65 & 0.74 & 0.67 & 0.55 & 0.73 \\
\bottomrule
\end{tabular}%
}
\end{table}

Frozen embeddings remain close to fine-tuned models on Causal eQTL, suggesting
that some variant-level signal is already encoded in pretrained representations.
However, Pathogenic ClinVar shows a larger gap, particularly for HyenaDNA.
Together with the splice-site results, this indicates that frozen
representations are less reliable when decisions depend on subtle, local, or
functionally complex sequence changes. Thus, variant-effect prediction provides
an independent local-sensitivity test that is consistent with the splice-site
findings.

\noindent\textbf{\textit{RQ5: How Does Embedding Geometry Explain Accessibility?}}

We quantify representation geometry using silhouette score, 5-nearest-neighbor
consistency, linear separability, and Fisher separation. These diagnostics test
whether task labels form separable structure in frozen embedding space.

\begin{table}[t]
\centering
\scriptsize
\setlength{\tabcolsep}{2.0pt}
\renewcommand{\arraystretch}{0.95}
\caption{Quantitative embedding geometry using frozen mean-pooled representations.}
\label{tab:geometry}
\resizebox{\columnwidth}{!}{%
\begin{tabular}{llcccc}
\toprule
\rowcolor{headergray}
\textbf{Task} & \textbf{Model} & \textbf{Sil.} & \textbf{kNN} & \textbf{Lin.} & \textbf{Fisher} \\
\midrule
Promoter All & NT & 0.044 & 0.883 & 0.917 & 0.044 \\
Promoter All & DNA2 & 0.031 & 0.885 & 0.949 & 0.029 \\
H3 & NT & 0.032 & 0.789 & 0.814 & 0.035 \\
\rowcolor{localred}
Splice All & NT & -0.011 & 0.465 & 0.597 & 0.012 \\
\rowcolor{localred}
Splice All & DNA2 & -0.011 & 0.438 & 0.515 & 0.011 \\
\bottomrule
\end{tabular}%
}
\end{table}

The geometry results mirror the predictive results. Promoter All has high
nearest-neighbor consistency and linear separability, whereas Splice All has
negative silhouette and much weaker neighborhood consistency. This shows that
the splice gap is not only a classifier artifact: splice classes are poorly
organized in frozen mean-pooled embedding space.

\noindent\textbf{\textit{RQ6: When Is Frozen Probing a Practical Alternative to Fine-Tuning?}}

Finally, we compare frozen probing with reproduced DNABERT-2 LoRA fine-tuning
under the same split, preprocessing, sequence length, metric, and readout
setting. This section is the matched-protocol frozen-versus-adapted comparison;
broader fine-tuned values in earlier tables are treated as benchmark- or
model-reported reference scores.

\begin{table}[t]
\centering
\scriptsize
\setlength{\tabcolsep}{2.0pt}
\renewcommand{\arraystretch}{0.95}
\caption{Controlled DNABERT-2 frozen probing versus LoRA fine-tuning.}
\label{tab:controlled_ft}
\resizebox{\columnwidth}{!}{%
\begin{tabular}{lccccc}
\toprule
\rowcolor{headergray}
\textbf{Task} & \textbf{FT} & \textbf{FT time} & \textbf{Frozen} & \textbf{Frz. time} & \textbf{Speedup} \\
\midrule
Prom. TATA & 95.96 & 665.5 & 93.38 & 99 & 6.7$\times$ \\
Prom. All & 97.15 & 9961.7 & 94.78 & 1027 & 9.7$\times$ \\
Splice All & 95.27 & 2433.3 & 49.57 & 475 & 5.1$\times$ \\
Enhancer & 54.10 & 1075.4 & 59.52 & 257 & 4.2$\times$ \\
H3 & 78.95 & 1461.1 & 70.99 & 286 & 5.1$\times$ \\
\bottomrule
\end{tabular}%
}
\end{table}

Frozen probing offers substantial computational savings, but the benefit depends
on task type. For promoter tasks, frozen representations remain close to
fine-tuning while being several times faster. For \textit{Splice All}, however,
the frozen probe is much weaker despite the speedup. This confirms the main
conclusion: frozen genomic representations are efficient and useful for many
global tasks, but local mechanistic tasks still benefit strongly from encoder
adaptation.

\section{Conclusion}

We investigated when frozen genomic foundation-model representations are sufficient for downstream sequence classification. Across regulatory, epigenetic, promoter, splice-site, and variant-effect tasks, frozen embeddings provide strong and efficient features for many global or composition-driven tasks, but are less reliable for local, position-sensitive mechanisms. Layer-wise probing, readout sensitivity, in-silico mutagenesis, variant-effect prediction, and embedding geometry indicate that local biological information is partially present, but not always accessible through final pooled embeddings and lightweight probes. Overall, frozen representations are a practical compute-efficient baseline, while challenging local tasks, particularly splice-site and pathogenic-variant prediction, may still benefit from adaptation or task-specific representation extraction.
\section{Limitations}

Our study has several limitations. First, the evaluation is limited to classification-oriented genomic tasks; applications such as sequence generation, retrieval, annotation transfer, and structured variant interpretation may require different representational properties. Second, our frozen-embedding pipeline relies primarily on pooled sequence representations. Although appropriate for evaluating representation sufficiency, pooling may dilute localized signals important for splicing and variant-effect prediction. Consequently, some performance gaps may reflect readout limitations rather than a complete absence of relevant information in the pretrained encoder.

Third, we evaluate a limited set of probes: logistic regression, RBF-SVM, and shallow MLP classifiers. While these models span different levels of decoding capacity, specialized attention-based, token-level, or motif-aware probes may recover additional information from frozen representations. Fourth, our robustness analysis focuses on selected single-nucleotide substitutions and does not cover insertions, deletions, structural variants, or long-range regulatory interactions. Finally, our efficiency analysis reports runtime but does not include hardware-normalized memory, energy, or carbon measurements.

Despite these limitations, our results provide a systematic view of when frozen genomic foundation-model representations are sufficient, when nonlinear decoding improves accessibility, and when task-specific adaptation or specialized representation extraction remains beneficial.
\bibliography{references}
\clearpage
\onecolumn
\appendix

\renewcommand{\thefigure}{A\arabic{figure}}
\renewcommand{\thetable}{A\arabic{table}}
\setcounter{figure}{0}
\setcounter{table}{0}
\section{Summary of Datasets}
\label{sec:appendix}
Table~\ref{tab:dataset_summary} summarizes the datasets used throughout our experiments, including the number of training and test samples and the number of output classes. 

\begin{table}[!htb]
\centering
\caption{Summary of Genomic Datasets Used in Experiments ($|$C$|$ indicates number of classes in the dataset). }
\label{tab:dataset_summary}

\begin{tabular}{|c|l|p{1cm}|p{1cm}|l|}
\hline
\textbf{\#} & \textbf{Dataset Name} & \textbf{Train Samples} & \textbf{Test Samples} & \textbf{$|$C$|$} \\
\hline
1 & Human Ensembl Regulatory         & 231,348  & 57,713   & 3 \\
2 & Drosophila Enhancers Stark       & 5,184    & 1,730    & 2 \\
3 & Demo Coding vs Intergenomic Seqs & 75,000   & 25,000   & 2 \\
4 & Demo Human vs Worm               & 75,000   & 25,000   & 2 \\
5 & Human Enhancers Cohn             & 20,843   & 6,948    & 2 \\
6 & Human Enhancers Ensembl          & 123,872  & 30,972   & 2 \\
7 & Human OCR Ensembl                & 139,804  & 34,952   & 2 \\
8 & Human Non-TATA Ensembl           & 27,097   & 9,034    & 2 \\
9 & Mouse Enhancers Ensembl          & 968      & 242      & 2 \\
\hline
\end{tabular}

\end{table}

\section{Performance comparison on different tasks for the three classifiers}
\begin{figure*}[!htb]
\centering
\includegraphics[width=\textwidth]{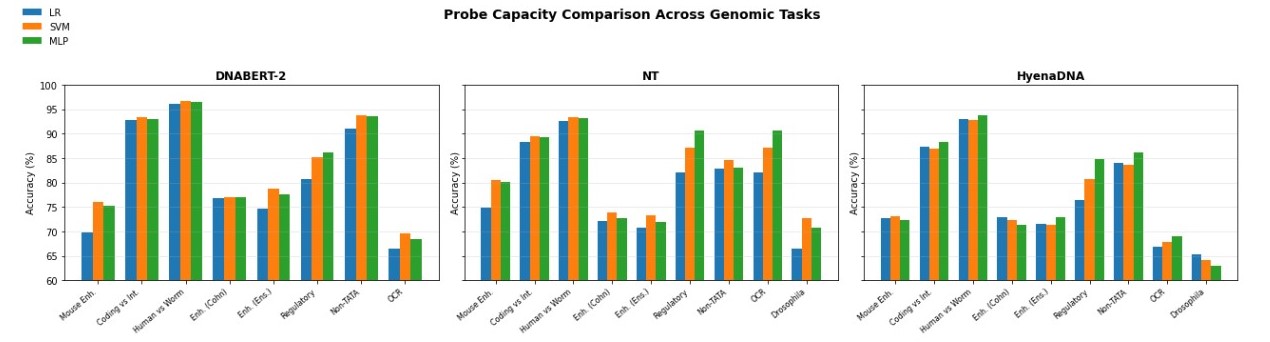}
\caption{Grouped bar comparison of probe models of increasing capacity (LR, SVM, MLP) across genomic classification tasks for DNABERT-2, the Nucleotide Transformer, and HyenaDNA. Each group corresponds to a dataset, with bars representing different probe models. Tasks with minimal variation across probes indicate linearly accessible representations, whereas tasks showing improvement with higher-capacity probes require nonlinear decoding. Tasks with consistently low performance across all probes suggest insufficient representation of task-specific signals.}
\label{fig:probe_capacity}
\end{figure*}

Figure~\ref{fig:probe_capacity} compares the performance of three probe models with increasing representational capacity—Logistic Regression (LR), RBF-SVM, and MLP—using frozen embeddings extracted from DNABERT-2, the Nucleotide Transformer, and HyenaDNA.

\section{Visualization of Embeddings}
\begin{figure}[!htb]
\centering
\includegraphics[width=\columnwidth]{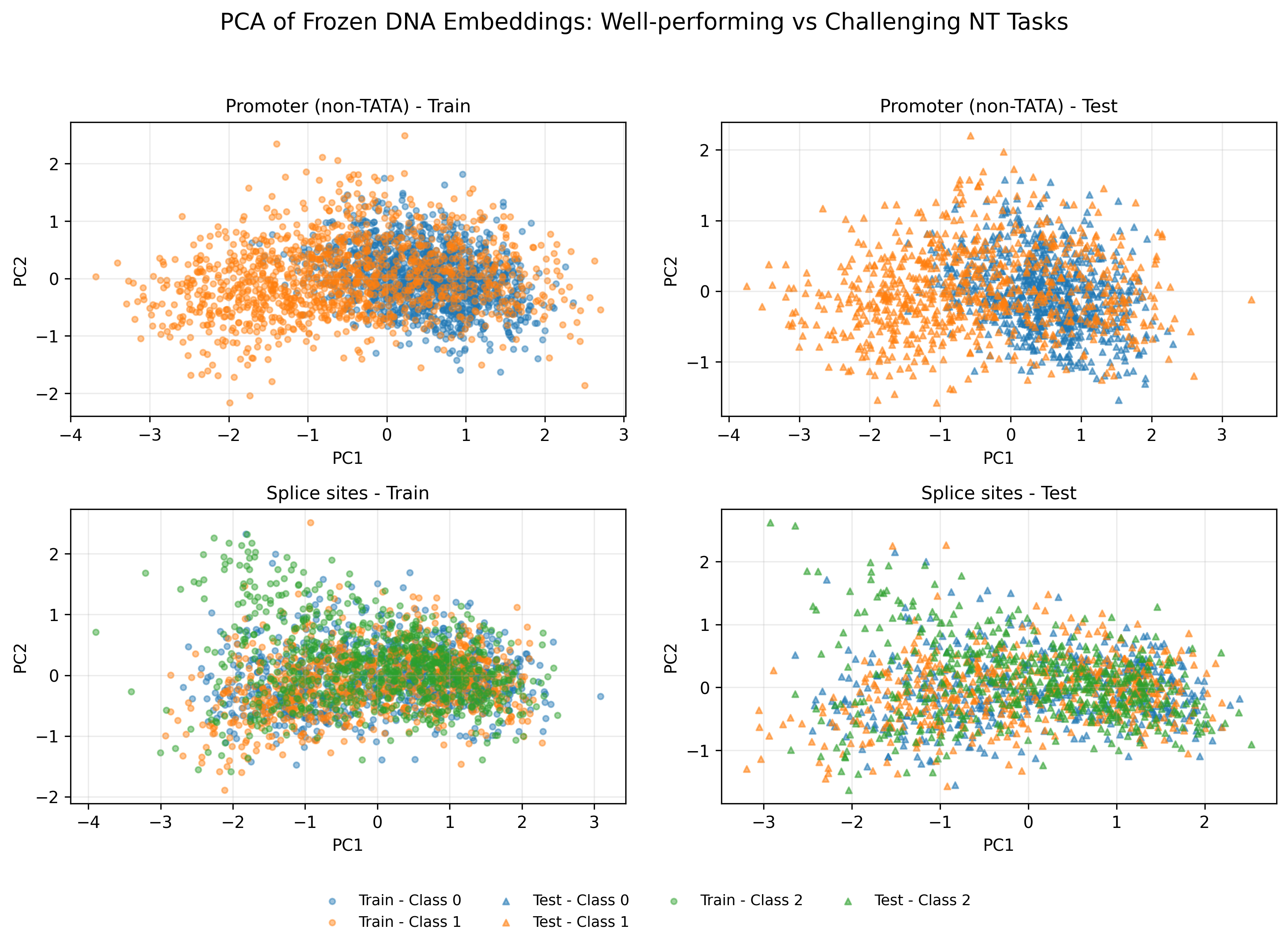}
\caption{PCA visualization of frozen DNABERT-2 embeddings for a well-performing task -promoter no tata and a challenging task -splice sites all from the NT benchmark. Colors indicate class labels, while marker shapes distinguish train and test splits. Promoter embeddings show strong train--test consistency despite limited low-dimensional class separation, whereas splice-site embeddings exhibit greater class overlap and dispersion, consistent with the reduced accessibility of fine-grained splicing signals.}
\label{fig:pca_good_hard}
\end{figure}

Figure~\ref{fig:pca_good_hard} presents two-dimensional PCA projections of frozen DNABERT-2 embeddings for a representative high-accessibility task (Promoter Non-TATA) and a challenging local task (Splice All), with colors denoting class labels and marker shapes distinguishing training and test samples.

\end{document}